\documentclass[letterpaper]{article} 
\usepackage{aaai2026}  
\usepackage{times}  
\usepackage{helvet}  
\usepackage{courier}  
\usepackage[hyphens]{url}  
\usepackage{graphicx} 
\usepackage{natbib}  
\usepackage{caption} 
\usepackage{algorithm}
\usepackage{algorithmic}
\usepackage{amsmath}
\usepackage{booktabs}
\usepackage{svg}
\usepackage{multirow}
\usepackage{newfloat}
\usepackage{listings}
\DeclareCaptionStyle{ruled}{labelfont=normalfont,labelsep=colon,strut=off} 
\floatstyle{ruled}
\newfloat{listing}{tb}{lst}{}
\floatname{listing}{Listing}
\begin{document}

\title{Via Score to Performance: Efficient Human-Controllable Long Song Generation with Bar-Level Symbolic Notation}

\author{
    \textbf{Tongxi Wang\textsuperscript{1}} \enspace
    \textbf{Yang Yu\textsuperscript{1}} \enspace
    \textbf{Qing Wang\textsuperscript{1}} \enspace
    \textbf{Junlang Qian\textsuperscript{2}}\thanks{Corresponding author}
\\
    \textsuperscript{1}Southeast University, China\\
    \textsuperscript{2}Nanyang Technological University, Singapore \\
    \texttt{
    tongxi\_wang@seu.edu.cn}
    \\\texttt{junlang001@e.ntu.edu.sg} \quad\quad\quad
}

\maketitle

\begin{abstract}
Song generation is regarded as the most challenging problem in music AIGC; nonetheless, existing approaches have yet to fully overcome four persistent limitations: controllability, generalizability, perceptual quality, and duration. We argue that these shortcomings stem primarily from the prevailing paradigm of attempting to learn music theory directly from raw audio, a task that remains prohibitively difficult for current models.  To address this, we present  $\mathcal{B}$ar-level $\mathcal{A}$I $\mathcal{C}$omposing $\mathcal{H}$elper (\textbf{BACH}), the first model explicitly designed for song generation through human-editable symbolic scores. BACH introduces a tokenization strategy and a symbolic generative procedure tailored to hierarchical song structure. Consequently, it achieves substantial gains in the efficiency, duration, and perceptual quality of song generation. Experiments demonstrate that BACH, with a small model size, establishes a new SOTA among all publicly reported song generation systems, even surpassing commercial solutions such as Suno. Human evaluations further confirm its superiority across multiple subjective metrics.
\end{abstract}

\section{Introduction}
Song generation \cite{yuan2025yuescalingopenfoundation} is believed to be the most challenging in music AIGC (AI-generated content) \cite{donahue2023singsong,du2023bytecover3,Wang2024MeloTransATA}, since it needs to handle both vocals and instruments and requires multiple tracks to be clear, coordinated, and layered. Current approaches generally adopt generative audio modeling \cite{huang2018music,copet2023simple,défossez2022highfidelityneuralaudio}. These methods are typically two-stage: first, the music is encoded into discrete codebooks, and then an auto-regressive model is used to generate the sequence of audio tokens.

\begin{figure}[t!]
\centering
\includegraphics[width=\columnwidth]{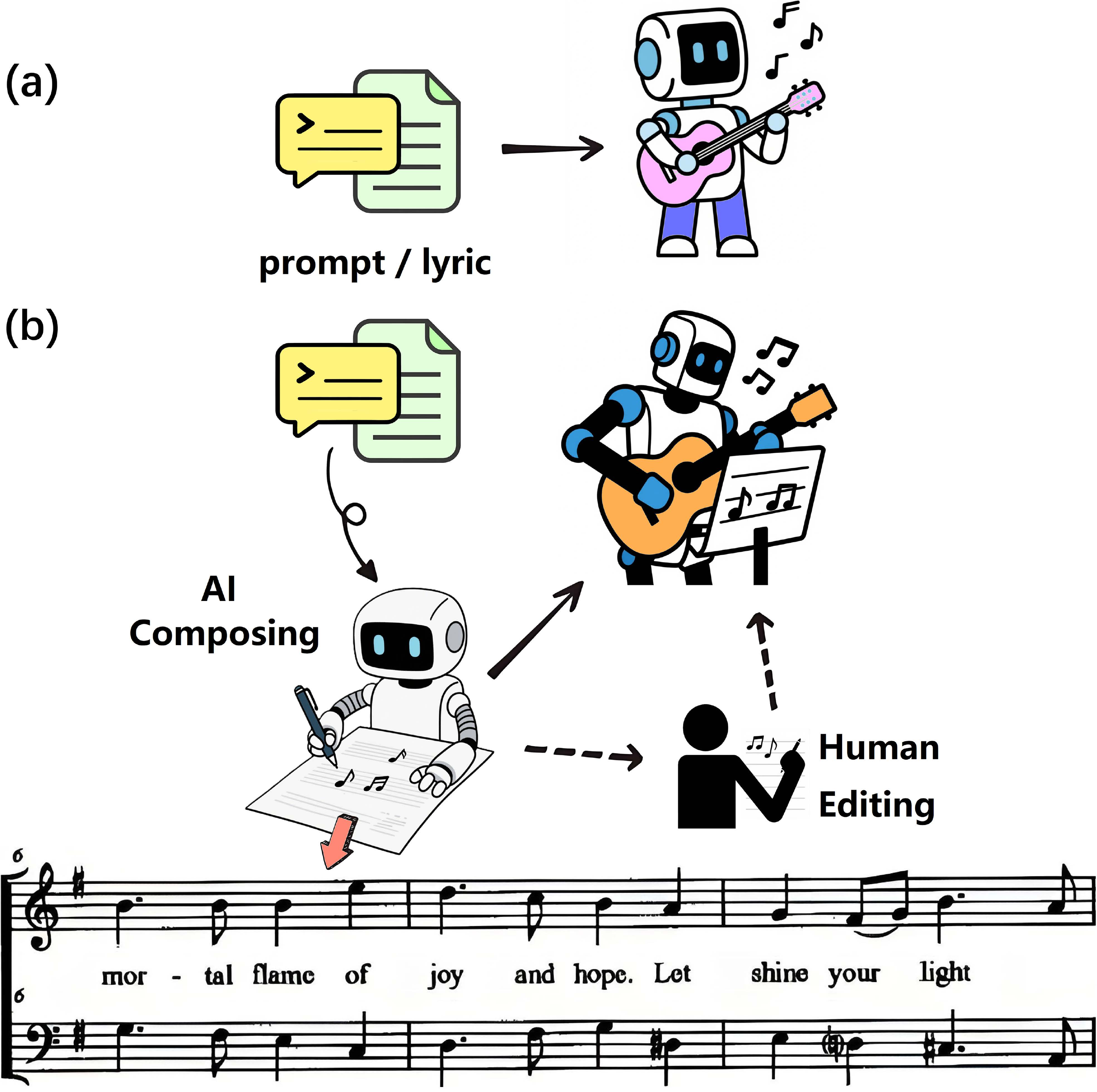}
\caption{
\textbf{(a)} Extemporaneous music performance by direct audio generation.
\textbf{(b)} Our via-score-to-performance approach: First, an AI composer writes a multi-track score aligned with the lyrics. Then, the song is rendered by specialized models for vocals and instruments. Users can edit the sheet music and lyrics for adjustments directly.
}
\label{fig:model-arch}
\end{figure}

However, we find these methods \textbf{not productive for users}. Our observations are as follows: (1) Lack of \textbf{general transfer ability}: Audio tokens used for different sounds are not interchangeable. Therefore, the zero-shot performance of these models would be weak. These models with audio tokens cannot well generate songs of new instruments or voices without modifying parameters\footnote{
Some studies \cite{Zhang2024MusicMagusZTA,Lv2023GETMusicGAA} are devoted to this problem, but more controllable solutions are needed.
}. 
The inflexible generation process may produce songs that are highly similar.
(2) Insufficient \textbf{human interpretability and editability}: The entire generation process is a black box. Users cannot easily edit details like rhythm or timbre. Multiple attempts are often needed for users to generate a satisfactory song, wasting creative computation and contradicting the human-centered vision of AIGC. 
(3) \textbf{Expensive computation}: Audio tokens contain excessive complexity, and generating a high-quality song requires too many tokens. Further, large music models are needed for bidirectional conversion between the discrete codebook and continuous signals.

We highlight an important fact: for any musician, \textbf{improvising while performing is far more challenging than composing first and then performing}. Previous generative audio models treat song generation as text-to-audio synthesis, requiring the model to simultaneously learn universal music theory, formal structure, composition, vocal techniques, accompaniment skills, and so on from raw audio. This task is overwhelmingly difficult, demanding massive data and broad coverage to achieve adequate generalization. To address this, we decouple the \textbf{learning of music theory} from the \textbf{mastery of music performance}. We generate scores aligned with lyrics, then render the score into vocals and multi-track instrumentals to form a complete song. By breaking the process into two phases, we simplify the task.

An alternative multimodal approach, symbolic text representation \cite{fang2025got,Wu2024MelodyT5AUA} has been applied in image and pure melody generation. In the form of plain text, symbolic representation naturally aligns with language models and often reduces computational cost. In audio generation, this idea is employed in sound-event \cite{mesaros2021sound} methods\footnote{
Some studies have also applied this paradigm to song generation \cite{Ding2024SongComposerALA,Liang2024ByteComposerAHA}.
}. However, these studies still treat songs as audio, neglecting the core organizing principle of music, leading to unsatisfactory auditory experience.

In this paper, we contend that \textbf{song generation is best grounded in musical scores} as it should be. This means treating the bar as the smallest semantic unit when structuring a song. The bar defines strong and weak beats and provides a framework for multi-voice and multi-track collaboration, ensuring rhythmic stability. 

\begin{figure}[ht]
\centering
\includegraphics[width=\columnwidth]{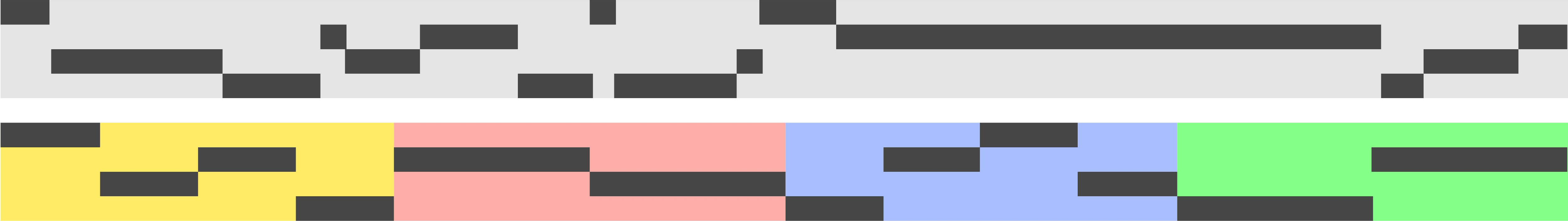}
\caption{
The upper track shows the layout from event-level generation, and the lower track shows the layout from bar-level generation. Music arranged by bar aligns better with music theory and is generally more pleasing.
}
\label{fig:eventVSbar}
\end{figure}

Building on this idea, we introduce \textbf{$\boldsymbol{\mathcal{B}}$ar-level $\boldsymbol{\mathcal{A}}$I $\boldsymbol{\mathcal{C}}$omposing $\boldsymbol{\mathcal{H}}$elper ($\boldsymbol{\mathcal{BACH}}$)}, the first model that brings structured scores to song generation. To prevent vocal and accompaniment streams from interfering with others, we introduce Dual-NTP in symbolic scores. Compared to prior work, BACH reduces both training and inference costs while enabling human edits at each stage. It generates songs in minutes, outperforming the best open-source system, which take hours. Against open-source baselines such as YuE and several closed-source commercial systems, including Suno, BACH sets a new state-of-the-art on composite metrics and also excels in human evaluations.

Our principal contributions are as follows:
\begin{itemize}
    \item We pioneeringly advocate a ``compose-first, perform-later'' strategy for song generation, with the bar serving as the primary semantic unit.
    \item We propose BACH, the first model built for bar-structured human-editable scores in song generation. Our model and code will be open-sourced.
    \item Extensive experiments show that, BACH generates SOTA pleasant songs, produces SOTA long durations, and demonstrates SOTA diversity, with SOTA speed and a small parameter count.
\end{itemize}

\section{Related Work}

\subsection{Music Generation and Singing Voice Synthesis}
Existing music-generation work spans a wide spectrum of approaches. These architectures are capable of processing diverse musical modalities—including Musical Instrument Digital Interface (MIDI), raw waveforms, and spectrograms—and demonstrate strong generative capacity in melody, harmony, and structure. 

Most prior audio generation efforts have focused on either instrumental music or a cappella vocals. In the instrumental domain, existing methods are typically limited to short segments (e.g., 10–30 seconds) and hindered by design choices that prevent simultaneous, high-quality vocal synthesis. Additionally, deep learning has greatly advanced singing-voice synthesis (SVS), offering fine-grained control over phonetic and timbral attributes. However, state-of-the-art SVS systems mainly generate unaccompanied vocals and rely heavily on explicit melodic supervision \cite{hong2023unisinger,liu2023audioldm,Chen2024ApplicationsAAA,Civit2022ASRA,Riedhammer2024ASOA}.


\begin{figure*}[ht]
\centering
\includegraphics[width=\textwidth]{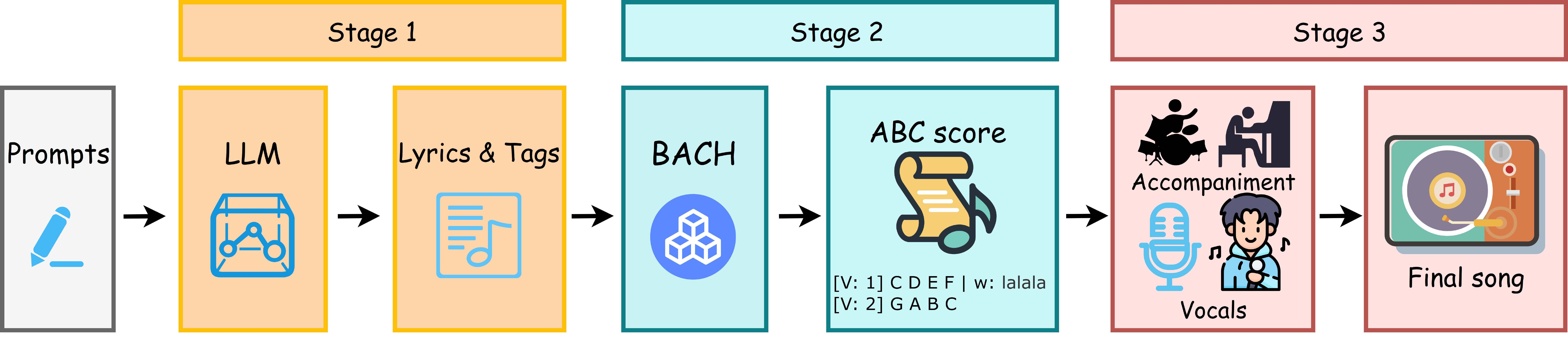}
\caption{
Our three-stage pipeline. BACH is in the second stage.
}
\label{fig:lyrics-tags}
\end{figure*}

\subsection{Song Generation}

Academic music-generation models still face significant challenges. Earlier works like MelodyLM  \cite{li2024accompanied}, SongGen  \cite{liu2025songgensinglestageautoregressive}, and SongCreator  \cite{lei2025songcreator} have difficulty maintaining coherence and high fidelity in musical excerpts longer than thirty seconds.

The lack of fully open-source implementations complicates reproducibility and further improvement. For example, Jukebox  \cite{dhariwal2020jukebox} uses a multi-scale Variational Quantized Variational Autoencoder (VQ-VAE), but it suffers from noticeable artifacts and limited controllability. Similarly, SongComposer \cite{Ding2024SongComposerALA} have introduced innovative Transformer-based architectures, yet still fall short of commercial systems in perceptual quality. Industry-developed systems have demonstrated impressive song-level generation capabilities \cite{bai2024seed}, though technical details remain undisclosed. Recently, the YuE model has come close to matching the performance of these proprietary systems in several areas, but its generation efficiency is still significantly lower.

\subsection{Symbolic Multimodal Generation}
ABC notation has proven to be an effective symbolic system in music generation \cite{yuan2024chatmusician,Qu2024MuPTAGA,Wu2023TunesFormerFIA}. However, this paradigm, textual multimodal generation via ABC notation, has not yet been applied to long-form song generation. BACH addresses this gap.

\section{Methodology}

\subsection{Background}
\textbf{ABC notation} \cite{walshaw2021abc} is a plain-text format that describes pitch, rhythm, bar-lines, lyrics, and multi-voice information.
\textbf{Bar-stream patching} \cite{wang2024exploringtokenizationmethodsmultitrack} has been shown to outperform BPE in music modeling, improving both perceptual quality and computational efficiency \cite{wang2025notagenadvancingmusicalitysymbolic}. The strategy converts an ABC score into a sequence of fixed-length, semantically aligned tokens.
\textbf{Dual-NTP} (Track-Decoupled Next-Token Prediction) \cite{yuan2025yuescalingopenfoundation} splits the originally entangled content into two separate token streams and has the same language model predict, at each time step, the next discrete token for both streams simultaneously. This preserves joint modeling while sharply reducing the risk that one track drowns out or interferes with the other \cite{qian-etal-2025-beyond}. We applied this idea and tailored it for the symbolic score of song generation. \\

\subsection{Our Pipeline}

Our main goal is to convert a brief user request into a complete song. The core of our approach is to align tokenization with the inherent harmonic and sectional structure of music while leveraging the text-processing strengths of language models. We adopt a ``compose first, play later'' paradigm, separating melody creation from audio rendering. This allows for a clearer presentation of the melodic structure of the song and improves both controllability and detail in the generated content.

We design a three-stage pipeline. First, a large language model parses the user prompt into multilingual lyrics with style tags. Second, these lyrics are passed into an ABC notation-based symbolic-score generation module, which segments the song into bars and separates vocal and accompaniment parts, producing a human-readable score. Finally, the vocal and accompaniment tracks are rendered separately and mixed to produce the complete song.

\begin{figure*}[t!]
\centering
\includegraphics[width=\textwidth]{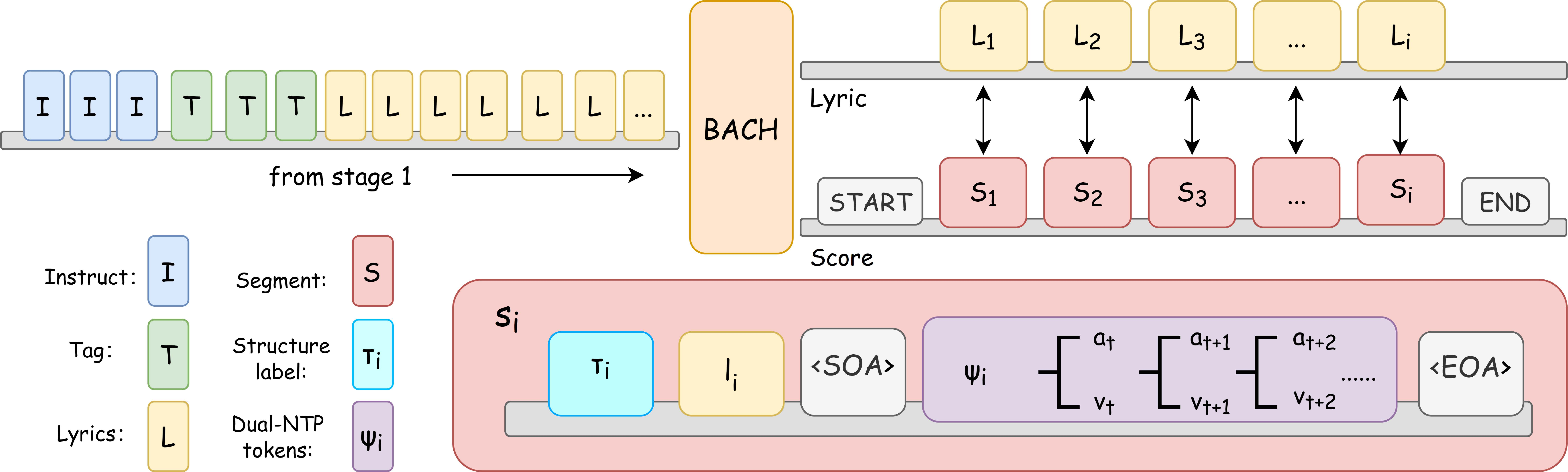}
\caption{
ABC lead-sheet text sequence generation method of BACH. Dual-NTP and Chain-of-Score are used to generate the ABC score.
}
\label{fig:tokenpair}
\end{figure*}

\subsection{Score-based Preprocessing and Postprocessing}

We have two components for preprocessing and postprocessing: \textbf{($1$) Lyrics \& Tags Generation} and \textbf{($2$) Audio Rendering}. Their functions are as follows: extract both the lyrics and their corresponding tags from the prompt of users, and convert the generated ABC notation into a complete song; So we can focus on how to transform the lyrics \& tags into ABC notation, which lies at the core of our work.

To improve multilingual capabilities, we employ the open-source large language model Qwen3.0  \cite{qwen3} to map user prompts into structured lyrics and style tags. Generated lyrics adhere to a hierarchical structure (intro, verse, chorus, bridge, outro) \cite{10248148}, with explicit section annotations; partial omissions are tolerated to maintain flexibility. This structural conditioning fosters a generation that balances stability and creativity, mitigating melodic monotony. Style tags are primarily drawn from a curated set of $200$ high-frequency tags (listed in the supplementary material), though additional tags can be added on demand, ensuring user control and stable generation.

After completing the conversion from lyrics and tags to ABC scores, the next step in our pipeline involves translating these structured outputs into a full musical composition. We first used the ABC-to-MIDI module to convert ABC notation into standard MIDI files for further processing. Then, FluidSynth was employed to turn the MIDI event stream into audio, generating the accompaniment. Simultaneously, VOCALOID was used to apply a deep neural network-based acoustic model to create the vocal track. The final result is achieved by mixing the two rendered outputs. This pipeline is fully automated, highly editable, and offers excellent scalability and control.

\subsection{$\boldsymbol{\mathcal{B}}$ar-level $\boldsymbol{\mathcal{A}}$I $\boldsymbol{\mathcal{C}}$omposing $\boldsymbol{\mathcal{H}}$elper ($\boldsymbol{\mathcal{BACH}}$)}

We introduce \textbf{BACH} ($\mathcal{B}$ar-level $\mathcal{A}$I $\mathcal{C}$omposing $\mathcal{H}$elper), the first model that incorporates both ABC notation and bar-stream patching into the song generation process. 

The normalized lyrics and style tags produced by the large language model (LLM) are fed into the module to generate the final song in ABC notation, aligned with the lyrics at the bar level. This module consists of two nested decoders: a patch-level decoder and a character-level decoder. Each patch is formed by concatenating the one-hot character vectors within the bar, which are then flattened and projected into a patch embedding via a linear layer. The patch-level decoder captures inter-patch temporal dependencies, and its final hidden state initializes the character-level decoder. This decoder then autoregressively predicts each character of the subsequent patch, producing the complete symbolic score in ABC notation.

\subsubsection{Bar-level Tokenization}

In bar-stream patching, the bar serves as the primary segmentation unit, in stark contrast to byte-level splits of Byte Pair Encoding (BPE). Each bar is divided into non-overlapping 16-character patches; any remainder is zero-padded. The resulting vocabulary consists of all possible 16-character ASCII strings, along with three special symbols\footnote{Table in appendix summarizes these symbols.}. Each patch token is subsequently routed to either the vocal or accompaniment stream within the Dual-NTP module.

\subsubsection{Symbolic Score Alignment Using Dual-NTP}
Popular LM-based approaches for modeling long random vector quantization (RVQ) sequences typically adopt a multi-stage design~ \citep{wang2023neural, 10158503, agostinelli2023musiclm}, where the first stage commonly uses a single codebook-0 token to represent each audio frame. Let 
\(\mathbf{x}_{1:T} = (x_1, x_2, \dots, x_T)\) 
represent a sequence of audio tokens, where each $x_t$ corresponds to one frame. In a NTP framework, the joint probability of $\mathbf{x}_{1:T}$ is factorized as: 
\begin{equation}
    p(\mathbf{x}_{1:T}) 
    \;=\; \prod_{t=1}^{T} p\bigl(x_{t}\mid x_{<t}; \,\theta\bigr),
    \label{eq:standard-ntp}
    \nonumber
\end{equation}
where $\theta$ represents the model parameters. This method works well for tokens $\mathbf{x}_{1:T}$ representing purely vocal or instrumental signals.

However, using a single token \(\,x_t\) for both vocal and accompaniment signals can cause the accompaniment to overshadow the vocals, reducing lyric intelligibility. To address this, we propose a method that splits each time step into two tokens: one for \textbf{vocal} and one for \textbf{accompaniment}. Dual-NTP demonstrate that emitting a discrete tuple $(v_t, a_t)$ at every time-step, where $v_t$ presents the vocal token and $a_t$ presents the accompaniment token.

To formally define this, let \(\mathbf{v}_{1:T} = (v_1, v_2, \dots, v_T)\) and \(\mathbf{a}_{1:T} = (a_1, a_2, \dots, a_T)\).  
In our method, each time step \(t\) outputs two tokens: \(v_t\) (vocal token) and \(a_t\) (accompaniment token).
The model’s sequence of tokens thus becomes:

\[
\bigl(\underbrace{v_1}_{\text{vocal}}, \underbrace{a_1}_{\text{accomp.}},\, 
\underbrace{v_2}_{\text{vocal}}, \underbrace{a_2}_{\text{accomp.}},\dots, 
\underbrace{v_T}_{\text{vocal}}, \underbrace{a_T}_{\text{accomp.}}\bigr).
\nonumber
\]

We factorize their joint probability as:

\[
    p\bigl(\mathbf{v}_{1:T},\, \mathbf{a}_{1:T}\bigr)
    \;=\;
    \prod_{t=1}^{T}\,
    p\Bigl(v_{t},\,a_{t}\;\Big|\;v_{<t},\,a_{<t};\,\theta\Bigr).
    \label{eq:dualNTP}
    \nonumber
\]

At inference time, the next pair \(\bigl(\hat{v}_t,\,\hat{a}_t\bigr)\) is chosen to maximize this joint conditional:

\[
    \bigl(\hat{v}_t,\,\hat{a}_t\bigr)
    \;=\;
    \arg\max_{(v_{t},\,a_{t})}\;
    p\Bigl(v_{t},\,a_{t}\;\Big|\;v_{<t},\,a_{<t};\,\theta\Bigr).
    \label{eq:dualNTPinfer}
    \nonumber
\]

making it straightforward to implement in standard AR decoding frameworks.

By maintaining the existing LM architecture, we take advantage of established pre-training infrastructures, ensuring seamless scalability. Separating vocal and accompaniment tokens allows for independent modeling, capturing subtle nuances, particularly in instrumentally complex sections. This also enables independent post-processing and mastering for each track.

\subsubsection{Chain-of-Score Structural Conditioning}
Traditional text-to-music (TTM) systems are limited to $\le$ $30$ s of context, whereas entire songs span several minutes. The all-in-one exploited the intrinsic sectional structure of music, decomposing songs into Intro, Verse, Chorus, Bridge, Outro, etc. To mitigate long-range degradation, we adopt and extend this idea to the lyrics in a symbolic music setting.

Within each structure section, text form segment labels, lyrics, and audio are paired together. From a full song perspective, structured text and audio tokens are \textbf{interleaved}. Special tokens are incorporated to indicate the start and end of the audio. Training samples are arranged as a serialized sequence, similar to the Chain of Thought in LLMs. Symbolic scores, lyrics, and other elements are concatenated into a single stream via Chain of Score. A training instance $\mathcal{D}_{\text{cos}}$ is serialized as:
\[
\mathcal{D}_{\text{cos}} = \text{Instruct} \circ \text{Tag} \circ \text{Lyrics} \circ \left( \bigcirc_{i=1}^{N} s_i \right) \circ \langle \texttt{EOD} \rangle,
\]
Specifically, $\circ$ denotes sequence concatenation. Instruct refers to the instruction, which serves as a task prefix. Tag refers to the musical tags, a style control string. Lyrics represent the raw lyric text provided before any segmented annotations. $\langle \texttt{EOD} \rangle$ is an end-of-document token.Besides, each segment $s_i$ is structured as follow:

\[
s_i = [\texttt{START}] \circ \tau_i \circ \ell_i \circ \langle \texttt{SOA} \rangle \circ \psi_i \circ \langle \texttt{EOA} \rangle \circ [\texttt{END}].
\]

$\tau_i$ is a structure label, $\ell_i$ representing lyric content of the segment, and $\psi_i$ denoting a sequence of Dual-NTP audio tokens, while $\langle \texttt{SOA} \rangle$ and $\langle \texttt{EOA} \rangle$ are tokens indicating the start and end of the score, they can be viewed as markers for audio refinement as well.

In summary, each document in CoS begins with an instruction, metadata, and raw lyrics, followed by a series of annotated segments, and ends with the $\langle \texttt{EOD} \rangle$ token.

By explicitly binding segment labels, segmented lyrics, and music patches, symbolic generation maintains song-level structural coherence and yields editable, interpretable inference traces.

\paragraph{In-context style transfer} We further equip the model with in-context learning (ICL) for stylistic adaptation. A reference excerpt about $30$ seconds is prepended:
\[
\mathcal{D}_{\text{icl}} = \mathcal{A}_{\text{ref}} \circ \mathcal{D}_{\text{cos}},
\]
with a delayed-activation strategy to prevent shortcut copying. Next, all patch tokens are de-tokenized: each token ID is mapped back to its 16-character ASCII string, right-padded spaces are removed, and control symbols are discarded. The characters are then concatenated in their original order and line-by-line to reconstruct the full ABC score.

\section{Experiment}

This section presents our experimental results. We begin with a brief overview of training strategy and data, followed by a description of the baselines and two evaluation suites used: (1) automatic music-generation metrics, and (2) human evaluation of the generated songs.

\subsection{Experiment Setup}

We combine Multitask Learning with Multiphase Training. For test-time strategies, we use forced decoding, restricting tokens to the audio scope until the model predicts \texttt{<EOA>}. For sampling and classifier-free guidance, the sampling hyperparameters are: top-k = 50, repetition penalty = 1.1, top-p = 0.93, temperature = 1, and max new tokens = 3000. Classifier-free guidance is applied with a scale of $s = 1.5$ to enhance the proportion of high-quality outputs.

\subsection{Data}
The dataset is derived from unpublished work, which is specially designed for our work. After tokenization, our dataset comprises 1B conditional vocal tokens, about $10$B unconditional music tokens (mixed and separated), and $2$B CoS music tokens. During annealing, $1$B CoS tokens were sampled from a high-quality subset and replicated four-fold to create a $4$B ICL dataset. Before annealing, the mixture ratio was conditional: unconditional = $3 : 1$ and music: speech = $10 : 1$. During annealing, only CoS and ICL data are used at a ratio of CoS: ICL = $2 : 1$. 

\subsection{Comparison Baselines}
At the time of writing, apart from YuE, no academic or open-source system offers usable long-form song generation. We therefore select four popular closed-source systems: Suno, Udio, Hailuo, and Tiangong. We reproduce YuE’s reported metrics and additionally train a lightweight version on our dataset, denoted YuE-light. Seven models are completely evaluated, including BACH-1B, YuE-7B, YuE-light-1B, Suno, Udio, Hailuo, and Tiangong. Each model produced 20 full-length English songs from diverse prompts. We kept the input prompts as consistent as possible. For models that only support lyrics-to-song and cannot handle prompt-to-song, we supplied the lyrics produced by our own Stage 1 as their input. Additionally, SongComposer, a recent entry, is also included in the comparison. Since we couldn't obtain a full reimplementation, we evaluated it on a subset of metrics using the limited examples released by the authors. We will use the name of our second-stage model BACH to refer to the entire song-generation system.

\subsection{Evaluation}

We conduct experiments on both model-based evaluation and human evaluation. 

For \textbf{model-based evaluation}, we adopt automatic metrics from three categories: distribution-matching (KL Divergence, FAD), content-aware (Audiobox-Aesthetic\footnote{Audio-aesthetics model from Meta, to assess whether the music sounds professional and pleasing.}: PQ, PC, CE, CU), and alignment (CLAP, CLaMP 3). KL Divergence and Frechet Audio Distance (FAD) gauge how closely the generated songs resemble real ones. CLAP and CLaMP 3 measure how faithfully the outputs follow the provided lyrics, style, and emotion prompts. Additionally, we report the vocal range spanned by the generated songs as a proxy for vocal agility, and we tabulate song duration to verify capability of the model in long-form generation.

For \textbf{human evaluation}, we conducted a study with $50$ raters: ten AI researchers and five trained musicians, while the remaining $35$ participants had no relevant professional background. None of them were involved in model training. Following prior studies \cite{yuan2024chatmusician,Latif2023SparksOLA}, we adopted an A/B test format. Evaluators blindly compared pairs of music pieces produced by two different systems. Comparisons on twelve criteria are listed in the supplementary material.

\begin{table*}[t]
    \centering
    \resizebox{\textwidth}{!}{%
        \begin{tabular}{r|cccccccc|c}
            \toprule
            \multirow{2}{*}{Model} & \multicolumn{2}{c}{Distribution Match} & \multicolumn{4}{c}{Content Based} & \multicolumn{2}{c}{Alignment} & \multirow{2}{*}{Overall} \\
            \cmidrule(lr){2-3} \cmidrule(lr){4-7} \cmidrule(lr){8-9}
            & KL $\downarrow$ & FAD $\downarrow$ & CE $\uparrow$ & CU $\uparrow$ & PC $\uparrow$ & PQ $\uparrow$ & CLAP $\uparrow$ & CLaMP 3 $\uparrow$ & \\
            \midrule
            SongComposer & -- & -- & 6.964 & 7.329 & 6.217 & 7.775 & 0.101 & 0.096 & -- \\
            Suno & 0.620 & 1.544 & \textbf{7.474} & 7.813 & 6.601 & 8.120 & 0.265 & 0.160 & 28.269 \\
            Tiangong & 0.708 & 2.547 & 7.421 & 7.766 & 6.060 & \textbf{8.220} & 0.244 & 0.114 & 26.570 \\
            Udio & 0.503 & \textbf{1.222} & 7.112 & 7.520 & 6.626 & 7.803 & \textbf{0.310} & 0.156 & 27.802 \\
            Hailuo & 0.756 & 2.080 & 7.350 & 7.737 & \textbf{6.793} & 8.132 & 0.265 & 0.106 & 27.547 \\
            YuE & \textbf{0.372} & 1.624 & 7.115 & 7.543 & 6.280 & 7.894 & 0.118 & 0.240 & 27.194 \\
            YuE-light & 0.423 & 1.604 & 7.097 & 7.333 & 6.129 & 7.897 & 0.152 & 0.193 & 26.774 \\
            BACH & 0.391 & 1.526 & 7.323 & \textbf{7.976} & 6.531 & \textbf{8.220} & 0.212 & \textbf{0.263} & \textbf{28.608} \\
            \bottomrule
            
    \end{tabular}
    }
    \caption{Automatic evaluation metrics across models. Bold numbers indicate the best result on each metric. The overall score is derived by summing all metrics. BACH achieves new SOTA on three individual metrics and the overall score.}
    \label{table:model-eval}
\end{table*}

\section{Results Analysis}

\subsection{Overall}
Table 1 summarizes the overall performance of our method versus seven baselines across all eight benchmarks. BACH sets a new SOTA on three individual benchmarks and on the overall composite score, despite using far fewer parameters and less training compute. Moreover, we conducted a comprehensive human evaluation. Results show that songs produced by BACH surpass the quality of the current best open-source model and most closed-source systems, and are competitive with the top performer, Suno. Our results are statistically significant, as confirmed by p-value testing.

\subsection{Model-Based Evaluation}
Table~\ref{table:model-eval} presents the automatic evaluation results, covering distribution-matching metrics (KL and FAD), Meta’s Audiobox aesthetic scores, and audio–text alignment scores (CLAP and CLaMP 3).

\textbf{Distribution-Matching Metrics.} We report KL and FAD to quantify how well generated audio matches the target distribution. BACH outperforms all commercial systems in terms of KL divergence by approximately 50\%, while ranking second in FAD. This demonstrates exceptional control of BACH over the generated content, effective audio quality, and distribution alignment. We attribute this to our ICL design and the structured decomposition of the song. 

\textbf{Content-Based Metrics.} Audiobox-Aesthetic scores above 7 indicate strong overall performance. BACH excels in Content-Based Metrics, achieving SOTA on two indicators and maintaining top scores overall. This highlights strong content quality and generation capability of BACH. We believe our contribution to Bar-level music score generation has significantly advanced our ability to produce harmonious and structured compositions.

\textbf{Alignment Metrics.} BACH achieved the highest alignment score under CLaMP 3 and surpassed the best existing open-source models on CLAP. Benefit from recent methodological improvements and a broader scale of pre-trained network resources, CLaMP 3 is more reliable \cite{yuan2025yuescalingopenfoundation} than CLAP scores. Strong performance of BACH on these metrics showcases the advantage of our ``compose first, play later'' strategy in terms of alignment.

\subsection{Human Evaluation}

\begin{figure}[ht]
\centering
\includegraphics[width=\columnwidth]{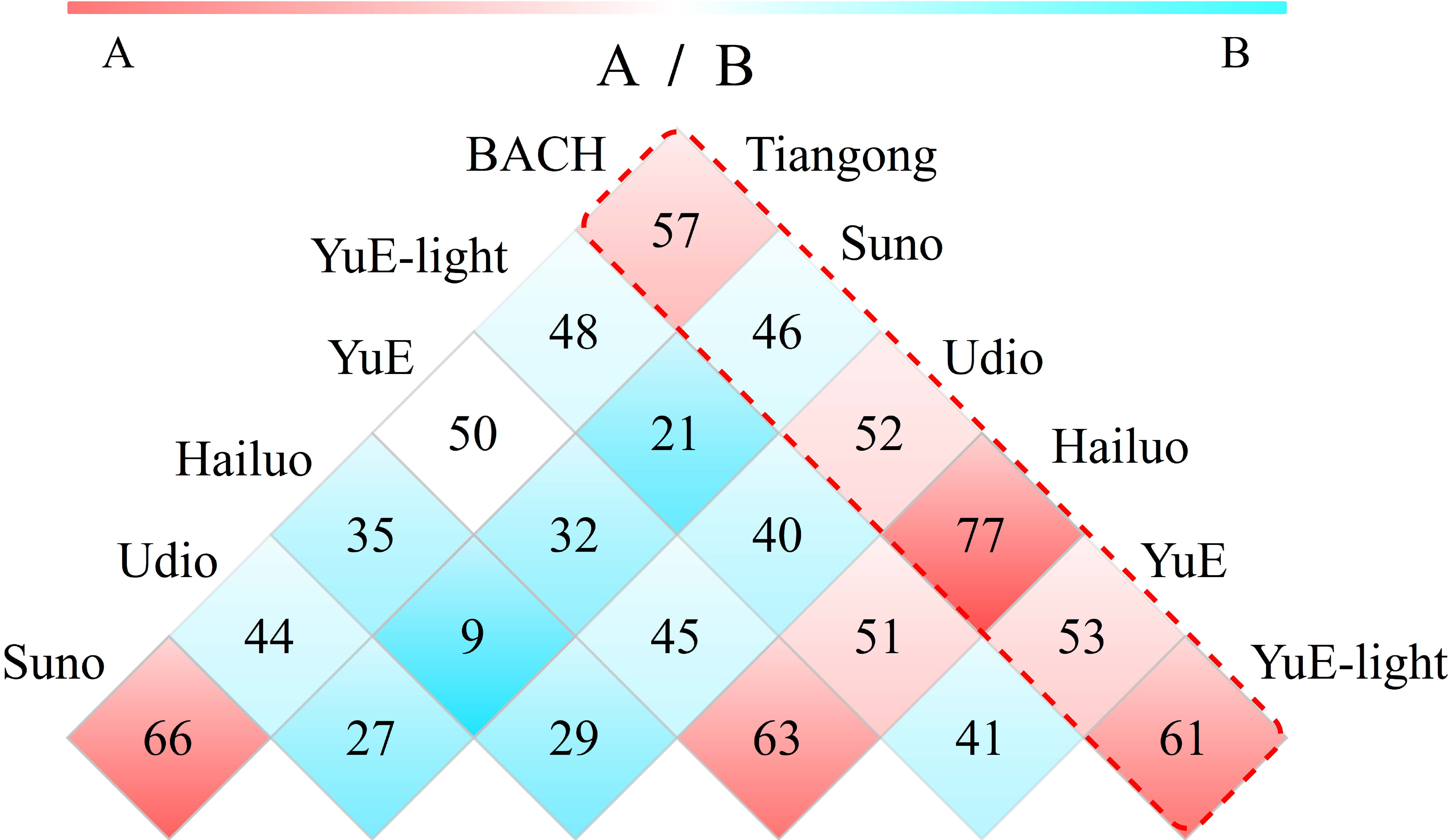}
\caption{Overall musicality heat-map (human A/B test). Colors and numbers denote the win rate of model A (left) against model B (right). Win rates of BACH are highlighted by the red box: it outperforms most models.}
\label{fig:overall-heatmap}
\end{figure}

\textbf{Overall Musicality (Fig.~\ref{fig:overall-heatmap}).} BACH achieves competitive mean preference and musicality versus six baselines: it significantly outperforms Hailuo, Tiangong, and YuE-light, moderately surpasses YuE and Udio, while still trailing the state-of-the-art Suno. These results indicate that top commercial products maintain an edge, yet BACH is a promising step toward high-quality open-source generation. 

\begin{figure}[ht]
\centering
\includegraphics[width=\columnwidth]{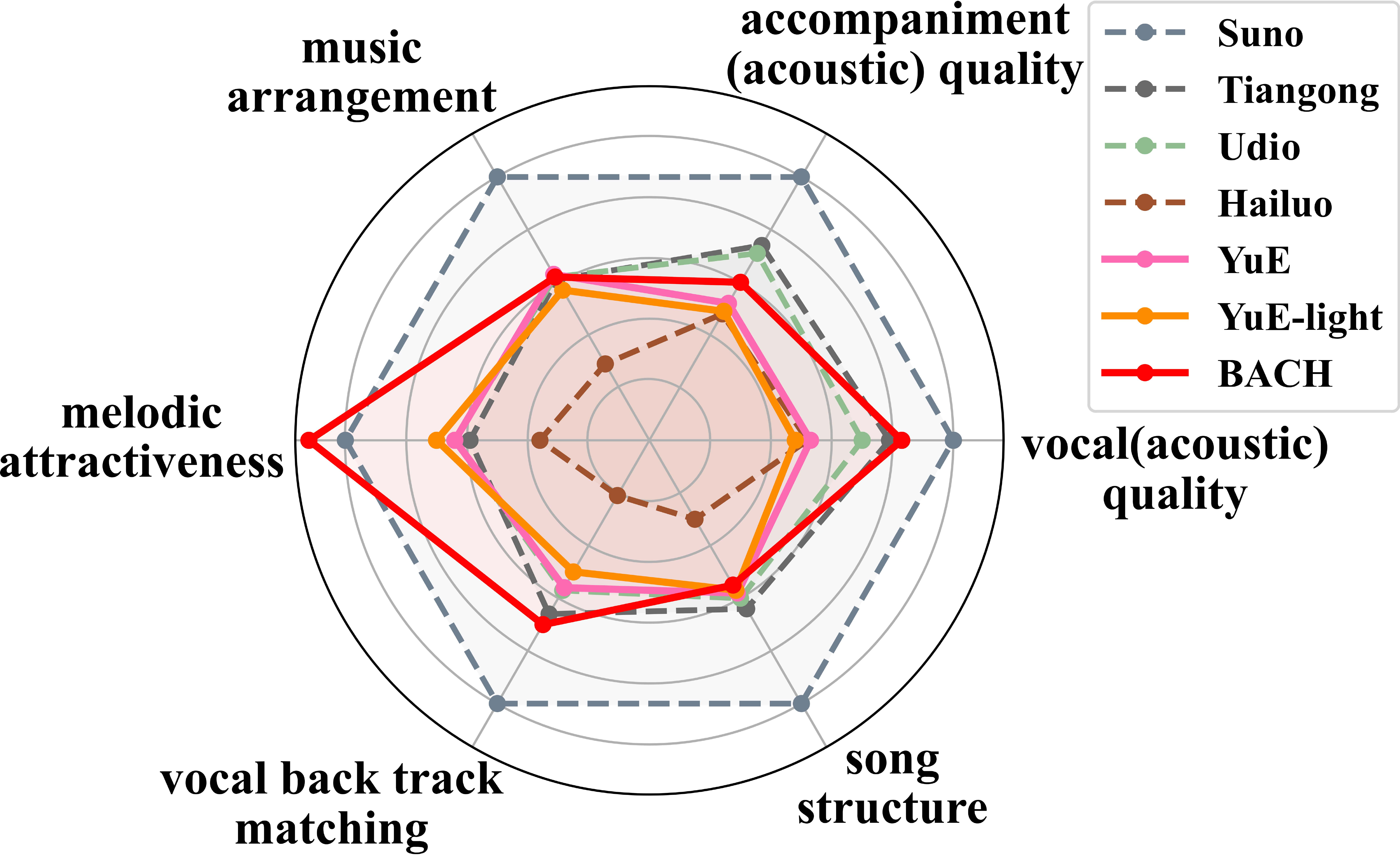}
\caption{Radar plot of musical quality dimensions. Despite its markedly smaller parameter count and training load, BACH delivers remarkably strong performance.}
\label{fig:quality-radar}
\end{figure}

\textbf{Aspects of Musical Quality (Fig.~\ref{fig:quality-radar}).} Radar plots show that BACH significantly outperforms YuE in vocal quality and melodic attractiveness, and shows improvements in accompaniment quality and vocal-backtrack matching. Compared to proprietary systems, BACH is on par with or superior to most in several areas, especially in vocal quality and melodic attractiveness, where it approaches or even surpasses Suno, validating our approach. This result highlights that the bar stream patching tokenization strategy greatly enhances melody generation and improves the overall performance across various aspects.

\begin{figure}[ht]
\centering
\includegraphics[width=\columnwidth]{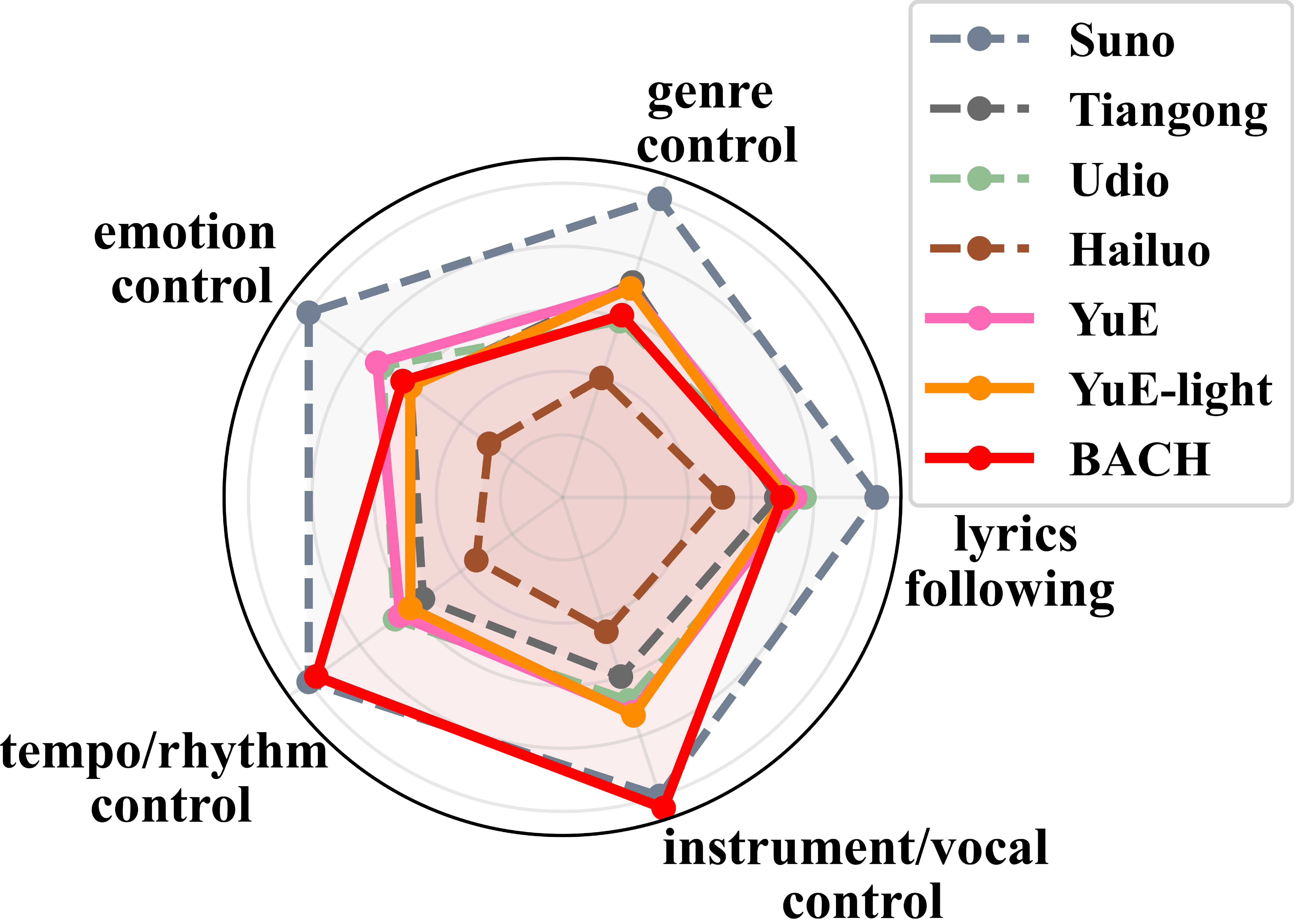}
\caption{Radar plot of controllability dimensions. Without human intervention, BACH excels in several aspects. }
\label{fig:controllability-radar}
\end{figure}

\textbf{Controllability (Fig.~\ref{fig:controllability-radar}).} Testing across five dimensions shows that BACH excels in tempo/rhythm and instrument/vocal control, highlighting the effectiveness of our alignment and rhythmic tagging. Genre and emotion control, however, are weaker, suggesting that direct symbolic generation may struggle with fine-grained steering. Post-score diffusion refinement could potentially address this limitation. These results not only validate our method but also suggest avenues for future improvement. 

\subsection{Vocal Agility}
\begin{figure}[ht]
\centering
\includegraphics[width=\columnwidth]{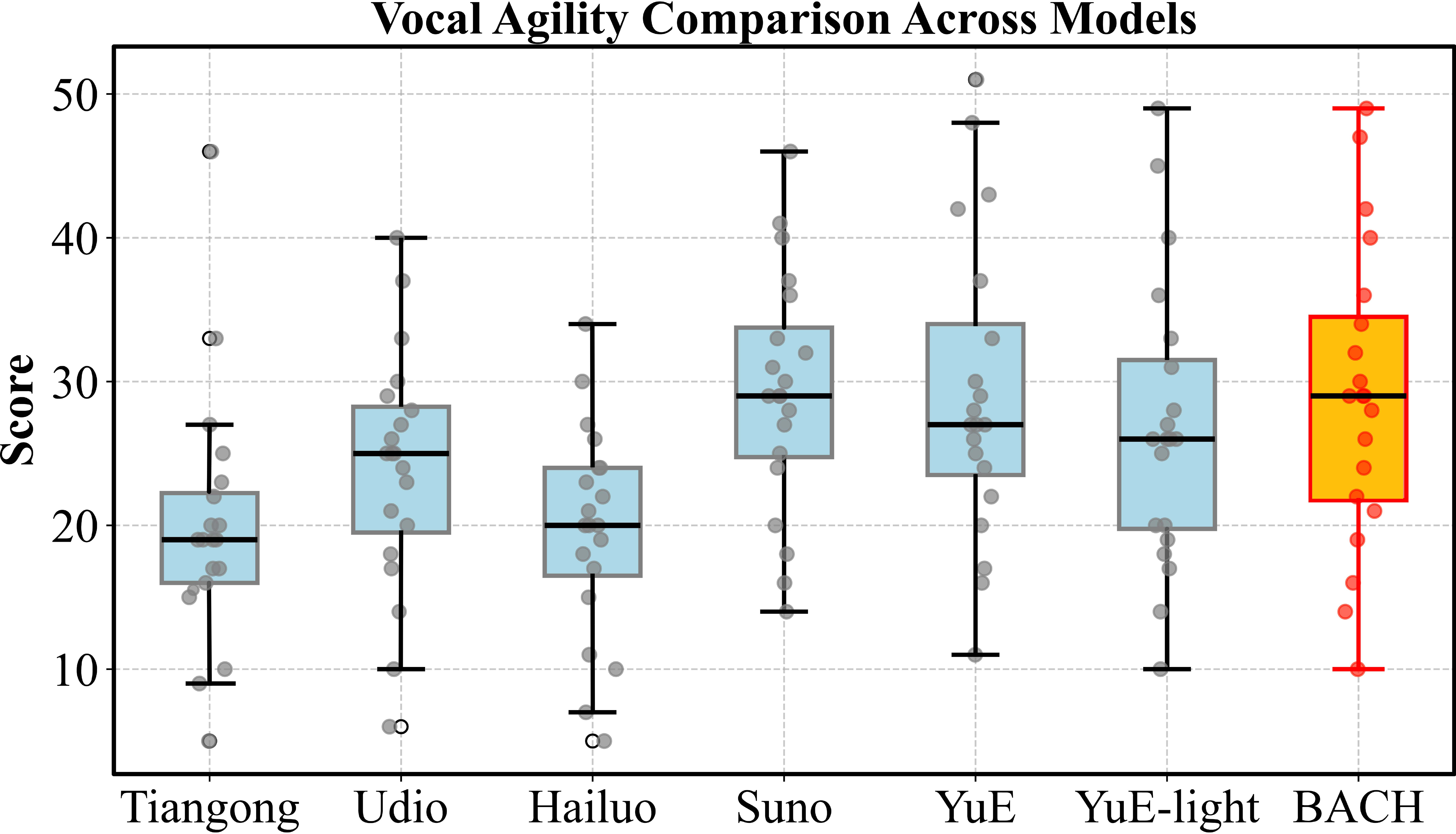}
\caption{Vocal agility boxplot. The y-axis indicates the vocal range spanned by each song; every dot represents one song. BACH achieves both the widest average vocal range and the highest upper bound efficiently.}
\label{fig:vocal-agility}
\end{figure}

As shown in the box plot in Figure~\ref{fig:vocal-agility}, the distribution of per-song vocal ranges reveals pronounced differences in vocal agility across systems; higher values indicate greater vocal expressiveness. Among all models, BACH exhibits one of the widest vocal ranges, on par with the best-performing closed-source system, Suno. By incorporating the ABC score, the model’s generalization ability was extended, enabling BACH to excel in generation diversity.

\subsection{Duration}
\begin{figure}[t]
\centering
\includegraphics[width=\columnwidth]{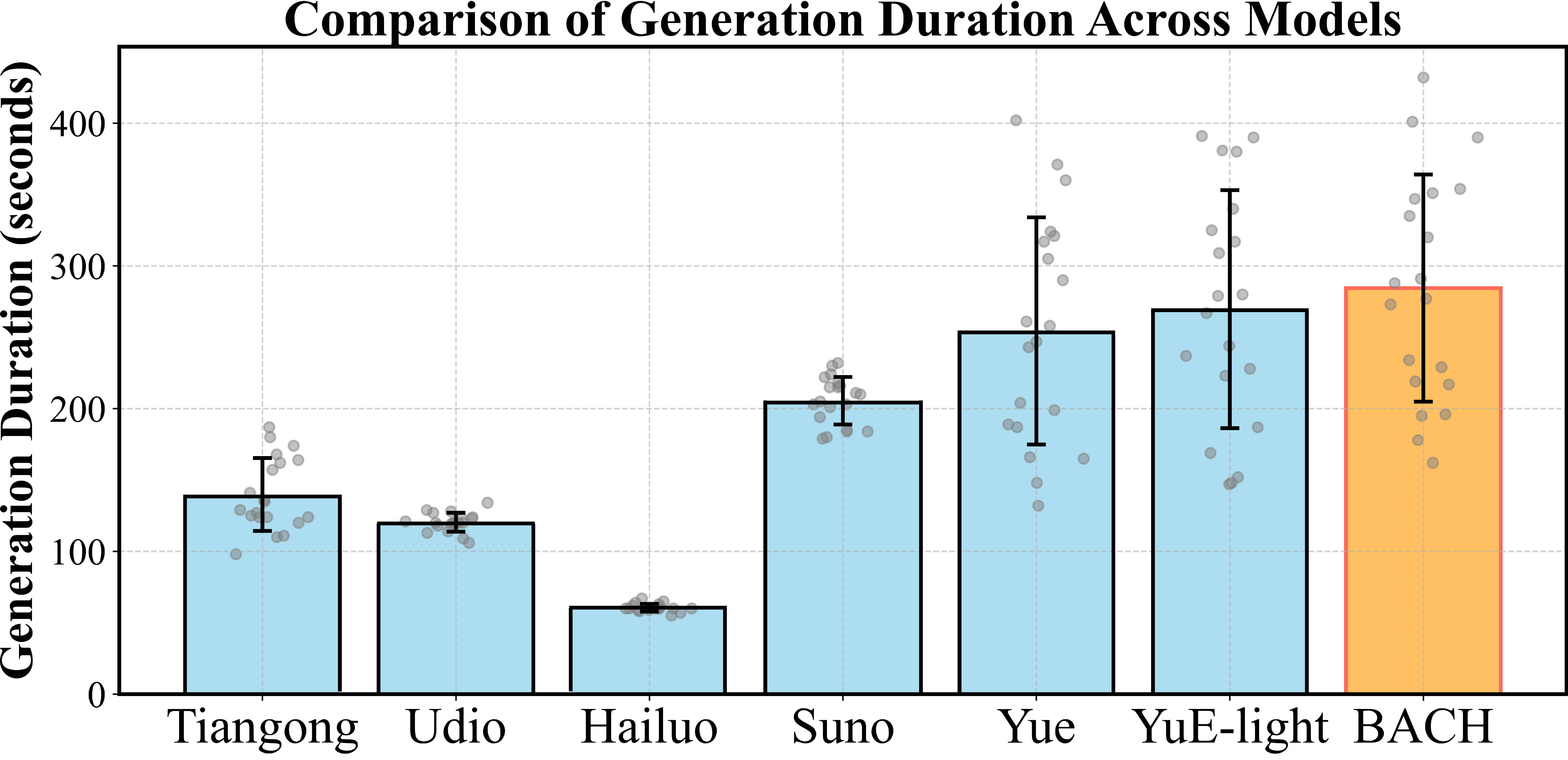}
\caption{Duration bar chart. The y-axis shows song duration in seconds; each dot represents a single song. BACH yields the longest average duration and the greatest upper-bound length.}
\label{fig:duration}
\end{figure}

Figure~\ref{fig:duration} shows the distribution of generated song durations, revealing significant disparities in length constraints. BACH produce the longest audio, with a much broader duration range than other models, demonstrating its ability to generate long-form songs. These results highlight our strength in modeling extended temporal dependencies, making it ideal for full-song generation.

\subsection{Efficiency}
Thanks to our ``compose first, play later'' design, we have not only optimized various performance metrics but also created a significant gap in computational resource consumption compared to YuE. At the same time, our minute-level generation efficiency offers a huge advantage within the scope of what we know. In summary, while achieving outstanding results like the ones mentioned above, we have also made notable contributions to optimizing computational resources.

\section{Discussion}

End-to-end generative modeling aims to map inputs directly to outputs using a single model, minimizing reliance on intermediate representations or manual feature engineering. This approach has seen notable success in modalities such as text and image, largely due to advances in autoregressive and diffusion models and the availability of large, high-quality datasets. However, the effectiveness of end-to-end modeling is closely tied to data abundance. In cases where data is both high-dimensional and relatively scarce—such as video—direct end-to-end modeling becomes less practical. For example, while video and text are both sequential in nature, a single video frame encodes far more information than a text token. End-to-end models operating directly on such high-dimensional data require substantial model capacity and, consequently, much larger datasets to avoid overfitting. Since large-scale, high-quality video datasets are much less common than text or image corpora, video generation often relies on hierarchical or multi-stage modeling strategies.

When increasing data volume is not feasible, reducing the dimensionality of model inputs and outputs is a common and effective strategy. Dimensionality reduction helps limit the number of model parameters, decreases computational demands, and can enhance generalization. This reduction can be achieved by discarding noise or redundant information within high-dimensional data.

There are two primary approaches for determining which information to retain or discard:
1. Learnable methods:  Machine learning models can be used to automatically discover low-dimensional representations that retain essential information, optimizing for downstream performance in tasks such as classification, generation, or reconstruction. Techniques such as Principal Component Analysis (PCA), autoencoders, and variational autoencoders (VAEs) fall under this category. These methods are particularly suitable when little prior knowledge exists or when the modality is complex.
2. Prior knowledge-based methods: In some cases, domain expertise and structural understanding of the modality can inform explicit dimensionality reduction strategies. For instance, in molecular generation, chemical bonds and functional groups are retained as essential information, while atomic vibration noise can be disregarded.

Music, as a modality, possesses strong domain priors and a relatively well-defined structure. Treating music generation as a general audio generation problem can unnecessarily increase complexity and dilute the benefits of domain-specific knowledge. By leveraging symbolic representations, hierarchical structures, and established musical conventions, the modeling process becomes more efficient and controllable. In our work, we demonstrate that incorporating domain-specific priors and explicit structure not only reduces model complexity but also improves both controllability and interpretability. This is supported by both objective and subjective evaluations, where our system performs on par with, or even surpasses, state-of-the-art commercial systems in several aspects.

Our findings emphasize the importance of tailoring modeling strategies to the unique properties of each modality rather than relying exclusively on end-to-end paradigms. Further progress in music generation may be achieved by exploring hybrid approaches that combine data-driven representation learning with domain knowledge. Additionally, expanding high-quality datasets and refining evaluation metrics to better capture human perception will be key to continued advancement in generative modeling for music and other complex modalities.

\section{Conclusion}
We present BACH, a long-form song-generation model, the first application of symbolic notation with bar-stream patching techniques to long-form song generation. Extensive comparisons with multiple baselines confirm the reliability of our approach. 

By introducing a new pipeline, we reduce computational overhead while matching or exceeding the performance of top commercial and open-source models. 

Our results on the chosen metrics are particularly noteworthy. BACH outperforms most models and approaches the level of Suno. By extending open-source models, we showcase the significant potential of bar-level music tokenization and symbolic score output in song generation.

In summary, BACH offers a novel strategy for long-form song generation, exhibiting strong performance and promising prospects in both efficiency and capability. 

\section{Limitation}

\textbf{Audio quality} can still be improved: during the neural music-rendering stage, we can further train a dedicated score-to-audio generation model that specializes in subtle note variations and unique vocal techniques. This enables the system to refine the details of the generated song and thereby enhance its overall quality.

\textbf{Data scarcity} remains a significant challenge, as there is a lack of sufficiently large open-source datasets in the industry. This limitation restricts the scale and diversity of training data. Expanding the volume and variety of publicly available datasets would likely lead to improved model performance.

\textbf{Inadequate evaluation metrics} present another limitation. The scientific rigor of evaluation metrics is crucial for the advancement of the field; however, existing automatic measures often deviate from human subjective perception. Relying solely on these metrics fails to fully capture a model’s true performance. Our system, like other song-generation approaches, still requires the integration of human evaluation. Therefore, the design of an assessment framework that can better capture underlying human preferences and judgments remains an urgent direction for the entire industry.

\bibliography{reference}

\end{document}